\documentclass{spie}
\usepackage{graphicx}
\usepackage{amsmath}
\usepackage{amssymb}

\DeclareMathOperator{\im}{Im}
\DeclareMathOperator{\re}{Re}
\DeclareMathOperator{\tr}{Tr}
\newcommand{\ham}{\mathcal{H}}
\newcommand{\curr}{{\mathcal{I}}}

\title{Low-frequency excitation of double quantum dots}

\author{Vyacheslavs Kashcheyevs
\skiplinehalf
Faculty of Physics and Mathematics, University of Latvia,
Ze\c{l}\c{l}u str. 8,  Riga, LV-1002, Latvia
\skiplinehalf
Institute for Solid State Physics, University of
Latvia, \c{K}engaraga str. 8, Riga, LV-1063, Latvia}

\authorinfo{E-mail: slava@latnet.lv}

\begin{document}
  \maketitle

\begin{abstract}
We address theoretically adiabatic regime of charge
transport for a model of two tunnel-coupled quantum dots connected in series.
The energy levels of the two dots are harmonically modulated by an external potential
with a constant phase shift between the two.
Motivated by recent experiments with
surface-acoustic-wave excitation, we consider two situations: (a) pure pumping
in the absence of external voltage (also at finite temperature),
and (b) adiabatic modulation of the current
driven by large external bias. In both cases we derive results consistent with published experimental data. For the case (b) we explicitly derive the adiabatic limit of Tien-Gordon formula for photon-assisted tunneling and compare it to the outcome of simple conductance modulation. A tutorial for adiabatic pumping current calculations
with the Green function formalism is included.
\end{abstract}


\section{INTRODUCTION}
\label{sec:intro}
Alternating external fields applied locally to a
confined electronic structure can result in a measurable time-averaged
current across the device. Generating dc current
by microwave excitation at frequencies of several GHz
and less is often referred to as \emph{pumping}.
Particular appeal of pumping both as a research tool and a potential source of applications for nanoelectronic systems is that low-frequency fields can be confined to wave-guides and
wires and thus conveniently delivered to samples at
a few Kelvin or lower temperatures.
This opens the way to explore pumping in situations when quantum coherence and single-electron charging effects matter.

Particular motivation for this work has been provided by two recent experiments \cite{MarkCNT08,Naber06} in which electrons in quantum dots have been subjected to an alternating piezoelectric potential of a running surface acoustic wave (SAW).
The experiment of Buitelaar \emph{et al.}\cite{MarkCNT08} focused on pure pumping current
with no external dc voltage bias.  Experimental results have been reported to be in a
good agreement with the theory of adiabatic quantum pumping\cite{Brouwer98} applied  to a simple two-level model (see definition  and discussion in Secs.~\ref{sec:Model} and \ref{sec:Adiabatic} of the present paper).
The experiment of Naber \emph{et al.}\cite{Naber06} has been designed to measure
the influence of an oscillating SAW potential on directed current driven by a constant external voltage but not necessarily by the SAW-induced potential itself. The results were found to be in a good agreement with Tien-Gordon formula\cite{tien1963,Stoof96} of photon-assisted tunneling.

In this paper we show that the results of these two experiments
are consistent within one and the same model: in the large bias limit,
adiabatic conductance modulation for the two-level system of Ref.~\cite{MarkCNT08}
reproduces the major features of the current traces  reported in Ref.~\cite{Naber06} (except for the fine structure due to non-adiabaticity). We derive analytically
the adiabatic limit of  Tien-Gordon formula and reproduce the criterion\cite{MoskaletsButtiker02} for the validity of an adiabatic approximation:
inverse life-time of the discrete charge states, $\Gamma/\hbar$, must
be larger than the modulation frequency $\omega$.

This paper has two major components. After introducing the model
in Sec.~\ref{sec:Model}, we discuss in detail the adiabatic pumping current
and its various limits (Sec.~\ref{sec:Adiabatic}). These results have been
used directly for interpretation of experimental data on pumping in carbon nanotube quantum dots\cite{MarkCNT08}. The discussion in Secs.~\ref{sec:Model} and \ref{sec:Adiabatic} is
more of a tutorial style. In the second part of the paper, Sec.~\ref{sec:NonAdiab}, we turn to the limit of large bias in which adiabatic conductance modulation dominates over  pure pumping. We compare predictions of the adiabatic time-domain expansion~\cite{Entin02form} (essentially, time-average of the Landauer formula) to the low-frequency limit of the photon-assisted tunneling through a double dot~\cite{tien1963,Stoof96}. This comparison uncovers precise agreement between the two theories (up to an overall integer  factor)
and establishes a rigorous adiabaticity criterion. Comparison to the experimental data of Naber~\emph{et al} identifies the adiabatic features in the measured current line-shape.
%
%
A brief summary and an outlook at open challenges in Sec.~\ref{sec:Conclusions} concludes the paper.

\section{THE MODEL AND ITS STATIC PROPERTIES}
\label{sec:Model}
The Hamiltonian $\ham$ describes two energy levels, $\varepsilon_1$ and
$\varepsilon_2$, with off-diagonal tunneling coupling $\Delta/2$,
connected to two external  reservoirs: level $\varepsilon_1$ to the left ($L$),
level $\varepsilon_2$ to the right ($R$). In the second quantized form,
\begin{align}
  \ham = \sum_{{i}=1,2}
  \varepsilon_{\alpha} d^{\dagger}_{\alpha} d_{\alpha}^{} +
   (d_{1}^{\dagger} d_{2}^{}+d_{2}^{\dagger} d_{1}^{}) (\Delta/2)
   + \sum_{k;{\alpha}=L,R} \varepsilon_{k {\alpha}} c_{k{\alpha}}^{\dagger} c_{k{\alpha}}^{}
   + \sum_{k} \left ( V_{k{L}}^{} c_{k{L}}^{\dagger} d_{{1}}^{} + V_{k{R}}^{} c_{k{R}}^{\dagger} d_{{2}}^{}+ \text{h.c.}
   \right )  \, . \label{eq:myHam}
\end{align}
The boundary conditions imposed onto $\ham$ are those of equilibrium reservoirs
at electrochemical potentials $\mu_L$ and $\mu_R$
(for the left and the right lead correspondingly) and at equal temperatures $T$.
The difference $\mu_L-\mu_R \equiv e V_{\text{bias}}$ specifies the external dc bias,
while the average $  (\mu_L+\mu_R)/2 \equiv \mu=0$ sets the reference level for energy.

We shall take $V_{k\alpha}$ in Eq.~\eqref{eq:myHam} to be independent of energy (the limit of a structureless, wide band ). In this case the only parameter characterizing the leads is the golden rule half-width $\Gamma_{\alpha} \equiv 2 \pi
\sum_k |V_{k{\alpha}}|^2 \delta(\mu_{\alpha}-\varepsilon_{k{\alpha}})$.
If no lead index is specified, the couplings will be assumed symmetric, $\Gamma_{L}=
\Gamma_{R} \equiv \Gamma$. When discussing
near-equillibrium properties (linear conductance, abiabatic pumping) we
shall assume vanishing bias conditions, $V_{\text{bias}}\to 0^{+}$.

The reason we omit physical spin index in Eq.~\eqref{eq:myHam}
is that the dots are assumed to be in the Coulomb blockade regime, so that double occupancy of a single dot in energetically prohibited. This assumptions of ``spinless \& non-interacting'' electrons can be viewed as a fermionic  representation of the $N \leftrightarrow N+1$ charge state transition
in an otherwise inert (that is, devoid of any internal dynamics) Coulomb-blockaded quantum dot. For a single dot this representation results in the well-known single resonant level (Breit-Wigner) approximation to Coulomb blockade~\cite{Alhassid00}.
Applicability of a non-interacting Hamiltonian like \eqref{eq:myHam}
to a pair of tunnel-coupled quantum dots is thoroughly discussed
in Sec.~V of a review\cite{Wiel03} by van der Wiel \emph{et al.}
While incomplete in its treatment of mutual capacitance and correlation effects,
the non-interacting charge-carrier approach adopted here
allows for a very detailed investigation of time-dependent effects because of the its fundamentally single-particle nature.

Some aspects of non-adiabatic pumping for the system defined by Eq.~\eqref{eq:myHam} have been considered recently \cite{Hanggi05} using a Floquet formulation for time-periodic non-equilibrium Green functions. Our model is also a special  (and previously unexplored) case of the a tight-binding model for SAW-induced adiabatic pumping as introduced and studied in Ref.~\cite{Aharony02PRL,VKAAOE03saw}\footnote{In the notation of Ref.~\cite{VKAAOE03saw}, one has to consider $N=2$, take the limit of $J \to \infty$ with $J_{\alpha}^2/J =\text{const}$ and identify $J_d$ with $\Delta/2$.}.

It is convenient\cite{VKAAOE03res} to define the retarded Green function matrix, $\hat{G}$, for the double-dot region in the following way\cite{Datta97},
\begin{gather}
  \hat{\Gamma}_{L}  = \begin{pmatrix}
                            \Gamma_{L} & 0 \\
                            0 & 0 \\
                          \end{pmatrix} \, , \quad
\hat{\Gamma}_{R}  = \begin{pmatrix}
                            0 & 0 \\
                            0 & \Gamma_{R} \\
                          \end{pmatrix} \, , \quad
 \hat{\ham}^{d}   = \begin{pmatrix}
                            \varepsilon_1 & \Delta/2 \\
                            \Delta/2 & \varepsilon_2 \\
                          \end{pmatrix} \, , \label{eq:myHamMatrix}\\
 \hat{G}(E)  = \left ( E-\hat{\ham}^{d}+ i \hat{\Gamma}_{L} /2 + i \hat{\Gamma}_{R} /2 \right )^{-1} \, .
\end{gather}
The transmission probability from left to right\cite{Datta97},
\begin{align} \label{eq:ourT}
  \mathcal{T}(E) & = \tr  \left [ \hat{G}^{\dagger} (E) \, \hat{\Gamma}_{R} \, \hat{G}(E) \, \hat{\Gamma}_{L} \right ] \, ,
\end{align}
determines the linear conductance $\mathcal{G}$ via (the finite-temperature) Landauer formula,
$ \mathcal{G} = (e^2/h) \int d E(- \partial f/\partial E) \mathcal{T}(E)$. Here and below
$f_{\alpha}(E) = \left [ 1+e^{(E-\mu_{\alpha})/k_B T} \right ]^{-1}$ is the Fermi distribution function.

For our system  the transmission probability is [see Eqs.~\eqref{eq:myHamMatrix} and \eqref{eq:ourT}]
\begin{align}
   \mathcal{T}(E)  = \frac{\Delta^2 \, \Gamma_L \, \Gamma_R}{4 \, D(E)} \, ,
\end{align}
 were the denominator $D(E) = |\det \hat{G}(E) |^{-2}$ is given explicitly by
\begin{align}
D(E) & \equiv \bigl([ E-\varepsilon_1]
   [E-\varepsilon_2] -\Delta^2/4 \bigr){}^2 +(E-\varepsilon_1)^2
\Gamma_R^2/4 + (E-\varepsilon_2)^2 \Gamma_L^2/4 + \Gamma_L \Gamma_R \Delta^2/8
    + \Gamma_L^2 \Gamma_R^2 /16 \, \\
     & =  ([E-E_1]^2+\Gamma^2/4)([E-E_2]^2+\Gamma^2/4) \quad \text{for } \Gamma_{L} = \Gamma_{R}  \, .
 \label{eq:Ddef}
\end{align}

\begin{figure}
\begin{center}
\includegraphics[width=6cm]{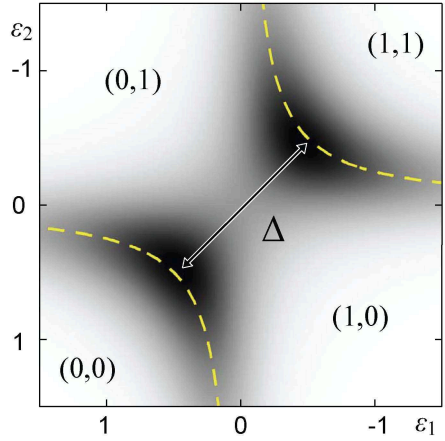}
\end{center}
  \caption{(Color online) Schematic plot of the stability diagram \cite{Wiel03}
  for our model as function of $\varepsilon_i$ (in units of $\Delta$). Dashed (yellow) lines show the eigenenergies $E_i$
  for no coupling to the leads ($\Gamma=0$). The tunneling splitting $\Delta$ is the maximum of $E_1-E_2$.
  The grayscale plot is the conductance $\mathcal{G}$ for $\Gamma/\Delta=0.2$, $k_B T/\Delta=0.1$, darker shade corresponds  to higher values of $\mathcal{G}$.
  The number in the parenthesis characterize the charge state for each of the dot.
   For $\Gamma/\Delta$ exceeding 1 the two dark blobs (triple points) merge into a quadruple point (not shown).\label{fig:stability}}.
\end{figure}

The eigenenergies $E_1$ and $E_2$,
   $ E_{1,2}=(\varepsilon_{1}+\varepsilon_{2} \pm \sqrt{(\varepsilon_1-\varepsilon_2)^2+\Delta^2 })/2  ,  $
can be understood as ``molecular orbital'' states of the double dot system.
For $\Gamma < \Delta$ the resonance lines of $E_{1,2}=0$ are resolved
in a two-dimensional plot of $\mathcal{T}(E=\mu)$ as a function of
$\varepsilon_1$ and $\varepsilon_2$, see  Fig.~\ref{fig:stability}.
Such diagrams are often referred to as stability diagrams\cite{Wiel03}
for systems of coupled quantum dots since they identify the regions
of different equilibrium occupation numbers for the dots.
Anti-crossing behavior seen in Fig.~\ref{fig:stability} is a hallmark of a tunnel-coupled double dot system and serves as a conceptual basis for charge-qubit approaches to quantum information processing~\cite{Fujisawa06}.

\section{ADIABATIC PUMPING}\label{sec:Adiabatic}
We envision pumping by modulating $\varepsilon_1$ and $\varepsilon_2$ with the help of a piezoelectric potential of a propagating SAW (or, perhaps, by direct gating). This requires calculating the response of the system to time-variation of the double-dot parameters\footnote{For simplicity, the effect of the modulating field on the tunnel couplings is assumed to be negligible.}.
One of the simplest scenarios~\cite{MarkCNT08} is a phase-shifted harmonic modulation:
  \begin{align} \label{eq:cont1}
  \varepsilon_1 (t)& = \varepsilon_0+\delta/2 +P \cos \omega t \, , \\
  \varepsilon_2 (t)& = \varepsilon_0-\delta/2 +P \cos ( \omega t +\varphi) \, . \label{eq:cont2}
\end{align}
where $\delta$ is a constant detuning between the levels and $P$ is the amplitude
of the external potential. However, in this section we shall focus on general properties of the adiabatic pumping current which are independent of  a particular choice of $\varepsilon_{1,2}(t)$.

Let us apply the general theory of adiabatic
transport in coherent strctures\cite{Entin02form}
to Hamiltonian \eqref{eq:myHam}  using a
Green function formalism\cite{VKAAOE03res}.
Instantaneous adiabatic current $\curr_{\alpha}^{\text{pump}} (t)$ entering lead $\alpha$
from the nanostructure is\cite{VKAAOE03res}
\begin{align} \label{eq:Ipump}
   \curr_{\alpha}^{\text{pump}} (t)& = \frac{e}{2 \pi } \int \frac{1}{2} \frac{-\partial [
   f_L(E)+f_R(E)]}{\partial E} \tr
   \left [ \hat{G} (E) \, \hat{\Gamma}_{\alpha} \, \hat{G}^{\dagger}(E) \, \frac{d}{d t} \hat{\ham}^{d} \right ] \,  d E \, .
\end{align}

The charge pumped by the system during one cycle of the periodic modulation with a circular frequency $\omega$ is \begin{gather}
   Q   \equiv
  e \int_0^{2 \pi /\omega} \!\!\!\!\!\!\!\! \left [ \curr_L^{\text{pump}}(t)-\curr_R^{\text{pump}}(t) \right ] \, dt = e \iint \frac{-\partial f(E)}{\partial E} \left [P_1(\varepsilon_1,\varepsilon_2,E) \, d \varepsilon_1
  + P_2(\varepsilon_1,\varepsilon_2,E) \, d \varepsilon_2
  \right ] d E \label{eq:currentGeneral} \, ,
\end{gather}
where
\begin{align}
 P_{1}(\varepsilon_1,\varepsilon_2,E)  &  \equiv \frac{- \Gamma_R \, \Delta^2+ \Gamma_L \,
\Gamma_R^2 + 4 \Gamma_L \, (\varepsilon_2-E)^2  }{ 16 \pi  D(E)} \, , \\
 P_{2}(\varepsilon_1,\varepsilon_2,E)  &  \equiv \frac{+ \Gamma_L \, \Delta^2- \Gamma_R \, \Gamma_L^2 - 4 \Gamma_R \, (\varepsilon_1-E)^2  }{ 16 \pi  D(E)} \, .
\end{align}
Since we have used a symmetrized current expression in Eq.~\eqref{eq:currentGeneral},
$P_1 \mapsto -P_2$ under $1 \leftrightarrow 2, L \leftrightarrow R$.

\begin{figure}
\begin{center}
\includegraphics[width=7.5cm]{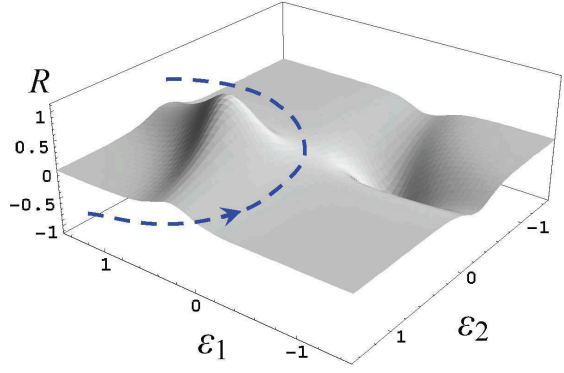}
  \caption{(Color online) The response function $R(\varepsilon_1,\varepsilon_2)$ \label{fig:figR}
  allows for determination of the charge pumped per period for an arbitrary contour.
  Blue dashed line shows an example of a pumping contour
  that produces the average current close to one electron per
  cycle.  The model parameters are $\Gamma/\Delta=0.2$, $k_B T/\Delta=0.1$; $\varepsilon_{1,2}$ are in units of $-\Delta$.
  For these values,   the maximum possible value for the pumped charge per period is $Q_{\text{max}}=0.981 e$.
  }
\end{center}
\end{figure}

As usual for two-parameter pumps\cite{Brouwer98},
it is convenient for the subsequent analysis to transform the contour integral in Eq.~\eqref{eq:currentGeneral} into an surface integral
using Green's theorem.
We write $Q$ as a double integral over the area
enclosed by the contour:
\begin{align}
  Q & = \pm e \int \!\!\!\int R(\varepsilon_1,\varepsilon_2) \, d \varepsilon_1 d \varepsilon_2  \, , \quad
  R  \equiv \int \frac{-\partial f(E)}{\partial E}  \left [ \frac{\partial P_2}{\partial \varepsilon_1}-\frac{\partial P_1}{\partial \varepsilon_2}
  \right ]d E \, . \label{eq:genQ}
  \end{align}
Volume under the two-dimensional plot of $R(\varepsilon_1,\varepsilon_2)$ above the pumping contour's image in the $R=0$ plane gives the value $Q$. The sign of $Q$ is determined by the direction of the contour [plus in Eq.~\eqref{eq:genQ} for counterclockwise direction].
This way of visualizing the pumping response is illustrated  in Fig.~\ref{fig:figR}.

Restricting the discussion to $\Gamma_{L}=
\Gamma_{R}$  allows for some further analytical progress.
Performing the integration over the energy $E$, we obtain explicitly
\begin{align}
 R (\varepsilon_1,\varepsilon_2) & = \frac{\Delta^2 \im \left [ S'(E_1)-S'(E_2) \right] }{2 \pi  (E_1-E_2)^3}
  - \frac{\Delta^2  \Gamma \left [
      (E_2-E_1) \re \{ S''(E_1)-S''(E_2) \}
        +\Gamma \im \{ S''(E_1)+S''(E_2) \}
  \right ]}{4 \pi (E_1-E_2)^2 [(E_1-E_2)^2 + \Gamma^2]}  ,
\end{align}
where $S(E) \equiv \Psi\left [ 1/2+(\Gamma+2 i E)/(4 \pi k_B T) \right ]$ and $\Psi$ is
the digamma function.

Let us consider also the maximal possible charge per period ($Q_{\text{max}}$) which is obtained when a large pumping contour completely encompasses the positive part of $R$.
For an arbitrary ratio $\Gamma/k_B T$ we find:
\begin{align} \label{eq:qmaxgen}
  Q_{\text{max}} & = e \int_{\Delta}^{\infty} \frac{\Delta^2
  \left \{ 2 (x^2+\Gamma^2) \im S(x/2)+ x \Gamma \left [x \re  S'(x/2)-\Gamma \im S'(x/2)
                          \right]
  \right \}
  }{\pi x^2 (x^2+\Gamma^2) \sqrt{x^2-\Delta^2}} dx \, .
  \end{align}
Specific limits of $R$ and $Q_{\text{max}}$ are illustrated below.
\begin{figure}
\begin{center}
\includegraphics[width=6cm]{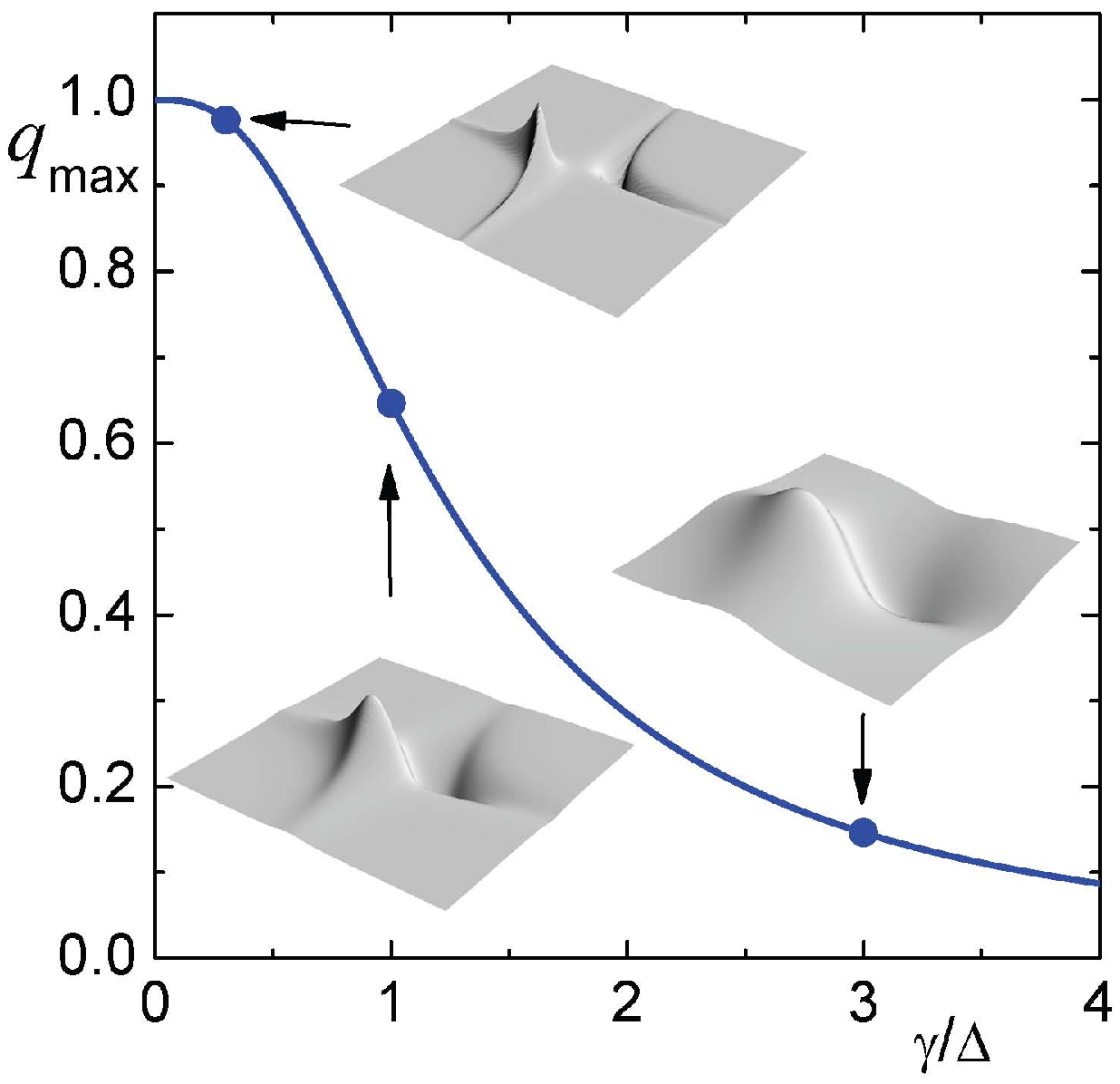} \hspace{0.25cm}
\includegraphics[width=6cm]{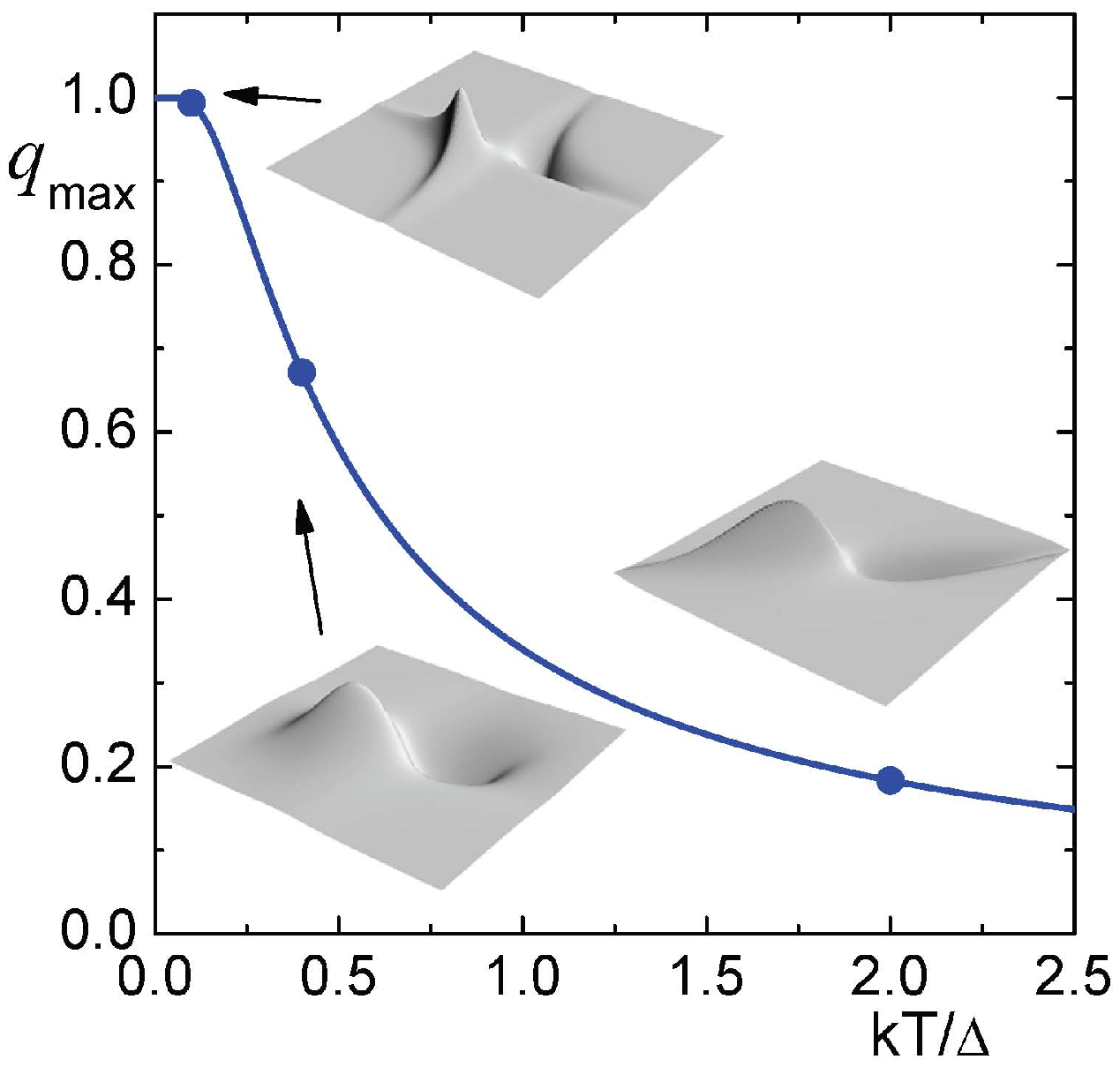}\\
\end{center}
\caption{(Color online) The maximal pumped charge per period in the limit of  $\Gamma/k_B T \to 0$ (left) and
   $k_B T/\Gamma \to 0$ (right)  as a function of $k_B T$ and $\Gamma$ respectively.
  For three representative ratios
  $\Gamma/\Delta =0.3$, $1$ and $3$ (left) and
  $k_B T/\Delta = 0.1$, $0.4$ and $2$ (right) the behavior
  of the response function $R$ is shown
  (absolute scale of $R$ varies from inset to inset).
  The values of $Q_\text{max}$ for the insets are
  ($0.976$, $0.646$, and $0.146$) and ($0.994$, $0.671$,  $0.183$)
  in the tunneling- and the temperature-dominated regimes  respectively. \label{fig:maxQ}
  }
\end{figure}

\subsection{Tunneling-dominated limit: $k_B T \ll \Gamma$}
The function $R(\varepsilon_1,\varepsilon_1)$ for $k_B T \ll \Gamma$ simplifies to
\begin{align}
  R(T\to0) = -\frac{\Gamma^3 \Delta^2 (\varepsilon_1+\varepsilon_2)}{8 \pi D^2(0)} \, .
\end{align}
The maximal possible charge \eqref{eq:qmaxgen} becomes
\begin{align}
   Q_{\text{max}}(T\to0)/e & = 1 - \left ( \frac{\Gamma}{\sqrt{\Gamma^2+\Delta^2}} \right )^3
 =       \begin{cases}
   1 -  (\Gamma/\Delta)^3 & \Gamma \ll \Delta \\
   3 \Delta^2/(2 \Gamma^2) &  \Gamma \gg \Delta
 \end{cases} \, .
\end{align}
This is illustrated in the left panel of Fig.~\ref{fig:maxQ}.
We see that charge quantization (meaning $Q \to e$) is possible
when $\Gamma \ll \Delta$. In this limit the two resonance lines
in Fig.~\ref{fig:stability} are well-defined, and pumped charge
quantization proceeds via the loading-unloading mechanism\cite{VKAAOE03res}.

\subsection{Temperature-dominated limit: $k_B T \gg \Gamma$}
In the opposite extreme of temperature-broadened resonance lines we have
\begin{align}
  R(\Gamma \to 0) = \frac{\Delta^2 [ f'(E_1)-f'(E_2)] }{2 (E_2-E_1)^3} \,
\end{align}
(prime denotes energy derivative). The corresponding maximal charge is
\begin{align}
 Q_{\text{max}}(\Gamma \to 0)/e & = \int_1^{\infty} \frac{\tanh [ x \Delta / ( 4 k_B T )]}{x^2 \sqrt{x^2-1}}
 d x
 =  \begin{cases}
   1 -\mathcal{O} (e^{-\Delta/(k_B T)} ) \, , & k_B T \ll \Delta \\
   \pi \Delta/(8 k_B T) \, , & k_B T \gg \Delta
 \end{cases} \, .
\end{align}
This limit of the model approximates the conditions of the seminal experimental
work of Pothier \emph{et al.}\cite{Pothier92}. Note that the ``exponential'' accuracy is possible only as long as $\Gamma \ll k_B T $ so that co-tunneling and other higher-order
processes remain below the level of thermal fluctuations.

\section{CONDUCTANCE MODULATION AT LARGE BIAS}
\label{sec:NonAdiab}
\subsection{Adiabatic pumping in the presence of external dc bias}
Let us turn to a situation when a constant
bias voltage $V_{\text{bias}}$ is applied together with a slow
modulation of the system parameters. General formalism for this scenario has been considered in Ref.~\cite{Entin02form}. The instantaneous adiabatic current $\curr({t})$ from left to right has been found to contain two additional components on top of the pure pumping contribution \eqref{eq:Ipump}:
\begin{align} \label{eq:Itotal}
   \curr(t) & = \curr^{\text{pump}}(t) +  \curr^{\text{mix}}(t) +
   \curr^{\text{bias}}(t) \, .
\end{align}
In our notation, the results of Entin-Wohlman \emph{et al.}\cite{Entin02form}
for the time-average current read
\begin{align} \label{eq:Ibias}
  \overline{\curr^{\text{bias}}(t)}  & = \frac{\omega}{2 \pi} \int_{0}^{2 \pi/\omega} \!\!\!\!\! d t \,
   \frac{e}{h}  \int \bigl [ f_{L}(E)-
f_{R}(E) \bigr ] \mathcal{T}(E) \, d E \, ,  \\
\overline{\curr^{\text{mix}}(t)}  &  =
\frac{\omega}{2 \pi} \int_{0}^{2 \pi/\omega} \!\!\!\!\! d t\,
   \frac{e}{h}  \int
   \frac{1}{2}
   \frac{\partial [ f_R(E)-f_L(E)]}{\partial E}
   \, \mathcal{T}(E) \frac{d
 \varphi_{\mathcal{T}}}{d t}  \, d E  \, . \label{eq:Imix}
\end{align}
Here $\varphi_{\mathcal{T}}$ is the overall transmission (Friedel) phase\cite{Langreth66}.

Interpretation of Eq.~\eqref{eq:Ibias} is straight-forward\cite{Entin02form} --- it is
the time-average of the current given by the finite-bias Landauer formula. Adiabatic perturbation of the system manifests itself in Eq.~\eqref{eq:Ibias} only as a parametric modulation of the transmission function $\mathcal{T}(E)$. The last remaining term,
Eq.~\eqref{eq:Imix}, is harder to interpret. It is non-zero only if both the bias $V_\text{bias} \not = 0$ and the adiabatic modulation are present at the same time (thus the notation $\curr^{\text{mix}}$).

We observe that both $\curr^{\text{pump}}$  [Eq.~\eqref{eq:Ipump}] and
$\curr^{\text{mix}}$  [Eq.~\eqref{eq:Imix}] contain the Fermi functions
$f_{\alpha}(E)$ only in the form of the respective energy derivatives.
If $ e V_{\text{bias}} \gg k_B T$ and the bias window $E \in [ \mu_R, \mu_L]$
is sufficiently wide to cover the whole energy range where the transmission function
$\mathcal{T}(E)$ is appreciable, then the conductance-modulation term, $\curr^{\text{bias}}$, dominates in the total current  \eqref{eq:Itotal}.
In the remaining part of this section we are going to explore the large bias limit
$ e V_{\text{bias}} \gg | \varepsilon_{1,2}(t)-\mu |, k_B T, \Gamma_{L,R} $
for the model defined in Eq.~\eqref{eq:myHam} and subjected to a harmonic modulation \eqref{eq:cont1} and \eqref{eq:cont2}.

\subsection{Large-bias current through a modulated double-dot system}
\begin{figure}
\begin{center}
\includegraphics[width=6cm]{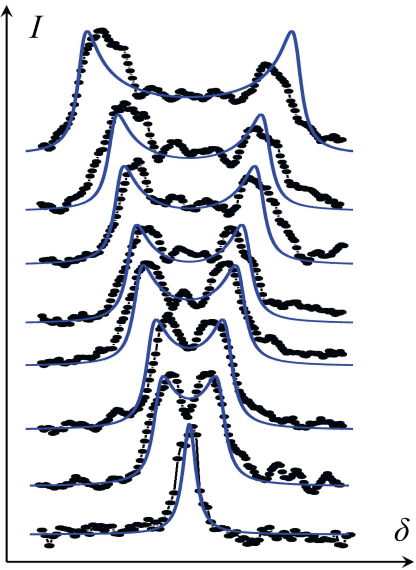}
\end{center}
  \caption{(Color online)
  Comparing the large bias limit of the adiabatic theory
  [Eq.~\eqref{eq:formtofit}, blue continuous lines] to the experimental data\cite{Naber06} (black circles) on a double quantum dot system formed electrostatically along an etched channel in a GaAs/AlGaAs heterostructure. The units are arbitrary (for quantitative details
  of the experiment, see the original work by Naber \emph{et al.}\cite{Naber06}).
  The horizontal axis is the difference between the gate voltages on the two
  dots (proportional to $\delta$), the vertical axis is the current measured in the presence of  a fixed bias voltage (same for all graphs) and a certain relative amplitude $P_{\text{eff}}$ due to applied SAW power. The latter is zero for the lowest trace and increases from bottom to top (individual traces are shifted along the $I$-axis for clarify). All graphs share the same scales for $\delta$ and for $P_{\text{eff}}$,
  but the overall magnitude of the current is fitted separately for each $P$.
  Note that the ripples between the broad peak are non-adiabatic features\cite{Naber06} that are adequately described by Tien-Gordon formula\cite{Stoof96}, see Eq.~\eqref{eq:TGbasic}. \label{Fig_comp_Naber}}
\end{figure}

The average dc current  in the large bias limit, $I^{\text{adiab}} = \overline{I^{\text{bias}}(t)}$,  is
\begin{align}
  I^{\text{adiab}}  = \frac{e}{h} \int \overline{\mathcal{T}(E)} d E
  & = \frac{\omega e}{h} \int_{0}^{2 \pi /\omega} \!\!
   \frac{\Delta^2 (\Gamma_L + \Gamma_R) d t }
{(\Gamma_L + \Gamma_R)^2 (1+\Delta^2/ [\Gamma_L \Gamma_R]) + 4 [\epsilon_1(t)-\varepsilon_2(t)]^2} \, . \label{eq:Ibiasadiab}
\end{align}
To derive the above equation we have performed the energy integration in
Eq.~\eqref{eq:Ibias} using Eq.~\eqref{eq:ourT} and extending the bias window to $\pm \infty$. The result for the time-averaged directed current \eqref{eq:Ibiasadiab}
is of the form
\begin{align}
   I^{\text{adiab}}  \sim \oint dt \frac{1}{A+ (\delta - P_{\text{eff}} \cos \omega t)^2} \label{eq:formtofit}
\end{align}
with $P_{\text{eff}} \equiv 2 P \sin \varphi /2$ and some positive constant $A$.
Thus the comparison to an experiment can be made without the detailed knowledge of
the tunnel couplings. Only the three scale factors --- for the current $I$, the detuning  $\delta$ and the modulation amplitude $P_{\text{eff}}$ ---  have to be adjusted.  We compare Eq.~\eqref{eq:formtofit} with the experimental results of Naber \emph{et al}
in Fig.~\ref{Fig_comp_Naber}.
Our result quantifies the qualitative adiabatic arguments of Naber \emph{et al.} concerning the increased time at resonance near the turning points of harmonic modulation\cite{Naber06}.

\subsection{Comparison to the non-adiabatic theory of Stoof and Nazarov}
Large bias limit for transport through two capacitatively coupled quantum dots connected in series has been considered theoretically by Stoof and Nazarov\cite{Stoof96}.
In the limit of a very weak inter-dot tunneling, $\Delta \ll \Gamma_L,
\Gamma_R, \hbar \omega, P_{\text{eff}}$, zero temperature and
strong asymmetry , $\Gamma_L \ll \Gamma_R$, they find the following
expression for the time-average current\footnote{This is equation (21) of Ref.~\cite{Stoof96} in our notation.}
\begin{align}
I^{\text{SN}} & =
 \frac{e \Delta^2 {\Gamma_R} }{ 4 \hbar}   \sum_{n=-\infty}^{+\infty}
  \frac{J_n^2(P_{\text{eff}}/\hbar \omega)}{\Gamma^2_R/4+(\delta - n \hbar \omega)^2} \, .
  \label{eq:TGbasic}
\end{align}
Here $J_n$ is the
$n$-th order Bessel function of the first kind.
A similar formula is known from the theory of photon-assisted tunneling between two superconductors due to Tien and Gorgon\cite{tien1963}, therefore
Eq.~\eqref{eq:TGbasic} is sometimes referred to by their name\cite{Wiel03}. It has been successfully applied to several experiments on microwave excitation of double quantum dots\cite{Wiel03}, as well as to the measurements of Naber \emph{et al.}\cite{Naber06} that we used in Fig.~\ref{Fig_comp_Naber}.
\begin{figure}
\begin{center}
\includegraphics[width=8cm]{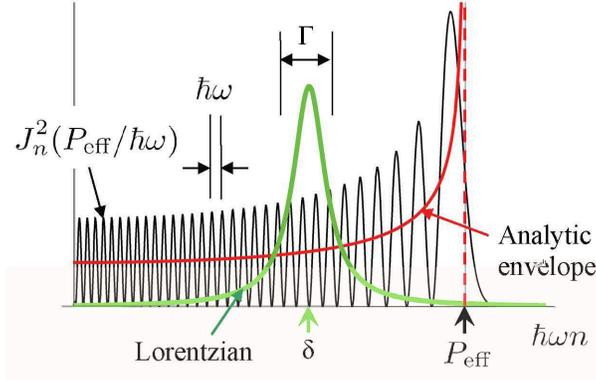}
\end{center}
  \caption{(Color online) Derivation of the adiabatic limit
  to the theory of Stoof and Nazarov [Tien-Gordon formula, Eq.~\eqref{eq:TGbasic}].
  \label{fig:BesselSketch}}
\end{figure}

We would like to compare Eq.~\eqref{eq:formtofit} to $\omega\to 0$ limit of Eq.~\eqref{eq:TGbasic}. This limit is singular and requires some care in implementation.
To this end, we note that Eq.~\eqref{eq:TGbasic} represents a convolution of the Bessel function with a Lorentzian, see Fig.~\ref{fig:BesselSketch}. Typical width of this Lorentzian along a (dimensionless) $n$-axis is $\Gamma_R/(\hbar \omega)$
while the period of oscillations of $J_n^2$ is of order one.
Evidently, in the limit $\Gamma_R \gg \hbar \omega$ the fine
oscillatory structure of  the current as a function of $\delta$ will be washed out.
This is the validity condition for the adiabatic limit to Eq.~\eqref{eq:TGbasic} that we are in a position to derive.

The Bessel function $J_n(z)$ oscillates for $z < n$ as~\footnote{The highly accurate approximation \eqref{eq:besselapprox}
is derived by taking $n=\xi z$ and applying the stationary phase expansion in $z^{-1}$
to an integral representation of the Bessel function,
$J_{n}(z) = (2 \pi)^{-1} \int_0^{2 \pi} \! \exp(i z \sin \varphi -i n \varphi) \, d\varphi$.}
\begin{align} \label{eq:besselapprox}
   J_{n}(z)\sim \frac{\cos \left [ z \, \theta (n/z) \right ]+ \sin \left [ z \, \theta (n/z) \right ]}{\sqrt{\pi} (z^2-n^2)^{1/4}} \, \text{ for }  |z|< n \, ,
   \text{ where } \theta(\xi)  \equiv \sqrt{1-\xi^2}-\xi \arccos \xi \, .
\end{align}
Equation~\eqref{eq:besselapprox} contains the analytic
form for the envelope of oscillations shown in Fig.~\ref{fig:BesselSketch}.
Replacing the sum in Eq.~\eqref{eq:TGbasic} with an integral, using Eq.~\eqref{eq:besselapprox} and dropping
the rapidly oscillating term gives
\begin{align}
  I^{SN} (\omega \to 0) & = \frac{e}{4 \pi  \hbar} \Delta^2 \Gamma_R \int_{-1}^{1} \frac{(1-\xi^2)^{-1/2} d \xi}
  {\gamma_r^2/4 + (\delta - \xi P_{\text{eff}}  )^2} =
   \frac{e}{4 h} \int_{0}^{2 \pi} \frac{ \Delta^2 \Gamma_R  \, d \tau}
  {\Gamma_R^2/4 + (\delta - P_{\text{eff}} \sin \tau)^2} \, .
\label{eq:TGapprox}
\end{align}
By derivation,  Eq.~\eqref{eq:TGapprox} approximates well the
full non-adiabatic result \eqref{eq:TGbasic} if $\Gamma_R \lesssim \hbar \omega$.

If we now expand the adiabatic result \eqref{eq:Ibiasadiab} to the leading order in $\Delta$ and take the limit $\Gamma_L \ll \Gamma_R$ then Eq.~\eqref{eq:TGapprox}
is recovered up to a factor of 4,
\begin{align}
  I^{\text{SN}}(\omega\to0) = 4 \, I^{\text{adiab}}.
\end{align}
To a reasonable degree of confidence we have excluded the possibility that the
factor of four arises due a notational or/and  algebraic mistake. Of course, we should bare in mind that the Hamiltonian of Stoof and Nazarov is different form ours: they consider the case of both strong intra-dot and inter-dot Coulomb repulsion. Only three states of an isolated double dot system are possible in their case while in ours there is the forth state --- the doubly occupied configuration $(1,1)$.

\section{CONCLUSIONS}
\label{sec:Conclusions}
Steady progress in experimental manipulation of nanoscale systems such as quantum dots and wires has recently reached a level enabling direct
comparison between theory and experiment for pumping in these devices. Many observed features of the pumping currents can be explained within a framework of single-electron models with time-dependent energy levels. In this paper we have looked at two examples of application for this framework to recent experiments. It is quite satisfactory to find agreement between different limits of different devices that comes within a single model. It is conceivable that in the future more refined  experiments will allow for quantitative characterization of the pumping current magnitude beyond the accuracy of a simple single-electron picture. Clear separation between the effects of discrete charging, quantum interference and many-body correlations in pumping through nanostructure seems posees future challenges both for theory and experiment.

\acknowledgments

The author is thankful to Bernd K\"{a}stner and Mark Buitelaar for continuous stimulating discussions on the subject. This research has been supported by the European Social Fund (agreement no.~2004/\-0001\-/VPD1\-/ESF\-/PIAA\-/04/NP/\-3.2.3.1/\-0001/\-0001/\-0063).


\end{document}